# Temperature dependence of interlayer coupling in perpendicular magnetic tunnel junctions with GdOx barriers

T. Newhouse-Illige[1], Y. H. Xu[1], Y. H. Liu[2], S. Huang[1], H. Kato[1], C. Bi[1],

M. Xu[1], B. J. LeRoy[1] and W. G. Wang[1,*]

1) Department of Physics, University of Arizona, Tucson, Arizona 85721, USA
2) Quantum Condensed Matter Division, Oak Ridge National Laboratory, Oak Ridge, Tennessee 37831, USA

Perpendicular magnetic tunnel junctions with $GdO_X$ tunneling barriers have shown a unique voltage controllable interlayer magnetic coupling effect. Here we investigate the quality of the $GdO_X$ barrier and the coupling mechanism in these junctions by examining the temperature dependence of the tunneling magnetoresistance and the interlayer coupling from room temperature down to 11 K. The barrier is shown to be of good quality with the spin independent conductance only contributing a small portion, 14%, to the total room temperature conductance, similar to $AlO_X$ and MgO barriers. The interlayer coupling, however, shows an anomalously strong temperature dependence including sign changes below 80 K. This non-trivial temperature dependence is not described by previous models of interlayer coupling and may be due to the large induced magnetic moment of the Gd ions in the barrier.

*wgwang@physics.arizona.edu

Spintronics[1,2], storing data with spin instead of charge, has been increasingly used for new computing devices. Taking advantage of the electron's spin degree of freedom in addition to its charge allows for non-volatile memory[3], logic[4], and sensor[5] devices with low energy requirements. The magnetic tunnel junction (MTJ)[6–10] is one of the most important functional units for these applications, where the information can be encoded in different resistive states. In general, MTJs can be switched with a magnetic field, a spin-polarized current by spin transfer torques (STT)[11,12], or a pure spin current by spin-orbit torques (SOT)[13,14]. It is also possible to change the state of an MTJ with voltage[15] for lower-power operation and easier integration with CMOS technologies. In MTJs with perpendicular magnetic anisotropy (pMTJ), the resistance can be switched through the combination of voltage controlled magnetic anisotropy (VCMA) and STT[16]. Subsequently, it has been demonstrated that pMTJs can be switched precessionally in a sub-nanosecond time scale by the effective in-plane field generated by the VCMA effect[17–19]. In addition, pMTJs can be manipulated by using voltage to control the interlayer coupling between the soft and hard ferromagnetic (FM) layers. This voltage controlled interlayer coupling (VCIC) effect was proposed theoretically[20–23] and demonstrated recently in pMTJs with a $GdO_X$ barrier, where the interlayer coupling field ($H_{IC}$) can be reversibly and deterministically controlled by voltage[24].

The nature of the interlayer coupling in $GdO_X$-pMTJs is of great interest itself. Generally, the observation of antiferromagnetic (AFM) coupling is more significant because a FM coupling could simply be due to the existence of pinholes in the barrier. A few models have been proposed to explain the coupling in MTJs with either in-plane magnetic anisotropy or perpendicular magnetic anisotropy (PMA). AFM interactions in an MTJ can be mediated by direct exchange coupling[25], described by wave function quantum interferences due to spin-dependent reflections at the FM/NM interfaces[26] or by the torques exerted by spin-polarized conduction electrons[27]. In MgO-based MTJs with large tunneling magnetoresistance (TMR), indirect coupling mediated through oxygen vacancies within the barrier has been shown to be responsible for the observed AFM coupling[28]. Finally, the AFM coupling in spin-valves and MTJs with PMA has been described by a roughness induced orange-peel magnetostatic coupling[29,30].

To evaluate the quality of the $GdO_X$ barrier and understand the interlayer coupling mechanism, we investigate the temperature dependence of $GdO_X$-based pMTJs from room temperature (RT) down to 11 K. The spin-independent tunneling contributes only a small portion to the total conductance, similar to AlOx and MgO based junctions, indicating a high quality barrier. More interestingly, $H_{IC}$ shows a strong temperature dependence, even changing sign below 80 K. This behavior has not been accounted for in existing models, suggesting that the interlayer coupling in $GdO_X$-pMTJs is strongly linked to the magnetic properties of the $GdO_X$ barrier.

Samples for this study were fabricated by magnetron sputtering in a 12 source UHV system with a base pressure of $10^{-9}$ Torr. The sample structure is $Si/SiO_2/Ta(8\text{ nm})/Ru(10\text{ nm})/Ta(7\text{ nm})/Co_{20}Fe_{60}B_{20}(0.7\text{-}0.9\text{ nm})/GdO_X(1\text{-}3.5\text{ nm})/Co_{20}Fe_{60}B_{20}(1.5\text{-}1.6\text{ nm})/Ta(7\text{ nm})/Ru(20\text{ nm})$. The layers were deposited under similar conditions as those used for fabricating MgO-pMTJs where TMR above 200% has been obtained at RT[31]. All layers were deposited with DC sputtering. The $GdO_X$ tunneling barrier was deposited by reactively sputtering of Gd target in Ar+$O_2$ atmosphere. After deposition, samples were patterned, using standard photolithography and ion etch procedures[32], into circular shapes with diameters (D) between 3 and 20 μm and a four wire contact geometry. Samples were measured on a custom probe station for RT measurements prior to wire bonding for temperature dependent measurements in an ARS sample-in-vapor cryostat with an external electromagnet. Although thick gadolinium oxide films exhibit a cubic $Gd_2O_3$ crystallization[33], the thin oxide layers in our MTJ samples are amorphous as revealed by the previous TEM study[24], and the oxidation state has not been precisely determined; therefore, the barrier is referred to here as $GdO_X$. The TEM study further revealed the interface between GdOX and CoFeB to be generally quite smooth, similar to what observed in AlOx/CoFeB and MgO/CoFeB junctions[24].

Ferromagnetic layers in contact with gadolinium oxide have been shown to exhibit many unique voltage controllable magnetic properties. Due to the large ionic mobility of oxygen in $GdO_X$, it has been shown that voltage applied across a $GdO_X$/FM interface can modify the speed of domain wall motion, the magnetic saturation ($M_s$) and the anisotropy field ($H_K$), even to the degree of changing the orientation of the magnetic anisotropy between in-plane and perpendicular[33–35]. GdOx was actually one of the first barriers used in the pioneering MTJ studies, where TMR was successfully observed at low temperature in junctions with in-plane magnetic anisotropy[36,37]. In the current study, room temperature TMR up to 15% was obtained in pMTJs with the use of an improved MTJ structure and CoFeB electrodes[24]. Representative TMR curves from these pMTJs at both RT (black) and 11 K (red) are displayed in Fig. 1, TMR being defined as ($R_{AP}$ – $R_P$)/$R_P$ where $R_{AP}$ and $R_P$ are the resistance in the antiparallel and parallel states, respectively. Both TMR curves show sharp resistance switchings and flat antiparallel plateaus as expected with pMTJs. Both the TMR and switching fields ($H_C$) increase considerably with the decrease in temperature as anticipated by previous results[8,10,38,39]. Additionally, both temperatures show a TMR curve symmetric about zero field as expected. Although bulk $Gd_2O_3$ shows antiferromagnetic ordering, its Neel temperature is below 4 K[40], lower than any of our testing conditions. All figures in this report correspond to a representative sample with a core structure of $Co_{20}Fe_{60}B_{20}$(0.85 nm)/$GdO_X$(2.5 nm)/$Co_{20}Fe_{60}B_{20}$(1.5 nm) except where noted. All other samples with different $GdO_X$ and/or CoFeB thicknesses show similar trends.

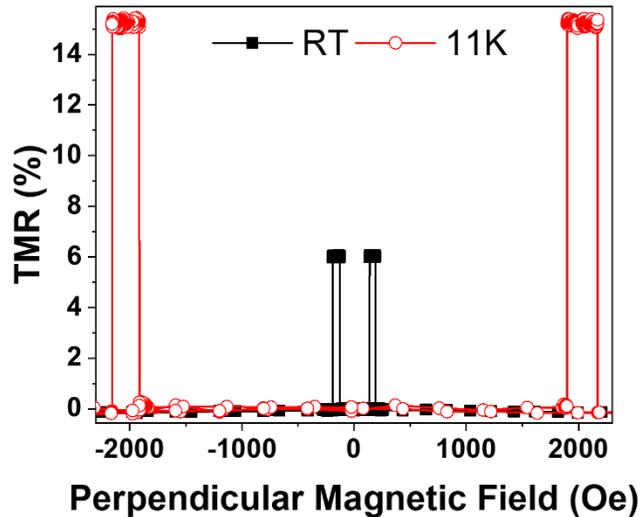

*FIG. 1. TMR curves of a representative sample at RT (solid black) and 11 K (open red).*

The increase in TMR at low temperature is illustrative of a tunneling mechanism with a spin-dependent contribution and a spin-independent contribution in the style of the two current model described by Shang et al.[41]. Following their model,

$$G(\theta) = G_T\{1 + P_1 P_2 \cos(\theta)\} + G_{SI} \quad (1)$$

where the first term is the spin-dependent tunneling and the second term, $G_{SI}$, is the spin-independent conductance. The temperature dependence of $G_T$, the direct elastic tunneling prefactor, is given by $G_T = G_0 CT/\sin(CT)$, with $G_0$ a constant and $C = 1.378 \times 10^{-4}\, d/\sqrt{\varphi}$, d being the barrier width in Å and φ the barrier height in eV. Rewriting equation (1) for symmetric FM electrodes, the standard TMR formula of Julliere[42] can be rewritten as $TMR = 2P^2/(1 + P^2 + G_{SI}/G_T)$, where P is the polarization of the CoFeB electrodes. The polarization is assumed to vary with temperature in the same way as the magnetization such that, $P = P_0(1 - \alpha T^{3/2})$.

Figure 2(a) shows that the conductivity of both the parallel (red) and antiparallel (black) states decreases with temperature. Looking at the conductance difference, $\Delta G = G_P - G_{AP}$, with $G_P$ ($G_{AP}$) being the conductance in the parallel (antiparallel) state, we can investigate only the spin-dependent conductance, since plugging $\theta = 0$ and $\pi$ into equation (1) and subtracting gives

$$\Delta G = 2 G_T P^2 = 2 G_0 P_0^2 (1 - \alpha T^{3/2})^2 CT/\sin(CT) \quad (2)$$

Fitting ΔG data to equation (2) gives $P_0$ and α equal to 0.521 and 6.30 x $10^{-5}$ K$^{-3/2}$, respectively. This value of α is in the expected range based on previous MTJs with MgO[43] and AlO$_X$ barriers[44].

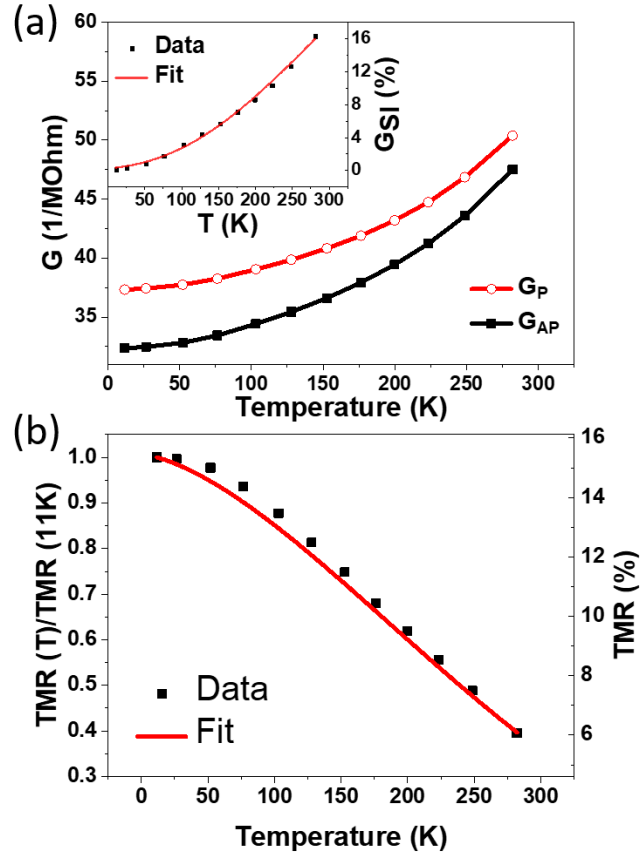

FIG. 2. (a) Conductance of the parallel (open red) and antiparallel (solid black) states. Inset. $G_{SI}$ normalized to $G_P$, data (points) and fit (solid line). (b) The temperature dependence of the TMR, data (points) and fit (solid line).

The inset of Fig. 2(a) shows the temperature dependence of $G_{SI}$. Assuming $G_{SI}$ is from conduction through localized defect hopping, which is mediated by coupling to phonons and so is strongly temperature dependent, we can model its temperature dependence. Under this assumption, $G_{SI}$ takes the form[45,46] $G_{SI} = \sum_N \sigma_N T^{(N-[2/(N+1)])}$, where N is the number of defect states involved in the hopping. The solid line in the inset of Fig. 2(a) shows that $G_{SI}$ is well fit by $\sigma_1 + \sigma_2 T^{1.33} + \sigma_3 T^{2.5}$ with $\sigma_1 = 6.2^{-8}$ $\Omega^{-1}$, $\sigma_2 = 1.1e^{-9}$ $\Omega^{-1}$ $K^{-1.33}$, and $\sigma_3 = 4.5e^{-12}$ $\Omega^{-1}$ $K^{-2.5}$. From this, it can be seen that the defect hopping at RT is strongly dominated by hopping through three impurity states. The quality of the barrier can be estimated by examining the proportion of the total conductance contributed by $G_{SI}$ at RT. The GdO$_X$-pMTJs investigated here have a $G_{SI}$ contribution of about 14% of the total conductance at RT. In MTJs where a large number of defect states were present in the barrier, much larger $G_{SI}$ was reported[47]. Here the $G_{SI}$ contribution in GdO$_X$ is similar to the range obtained with AlO$_X$[41,44] and MgO[43], 12-20%, indicating GdO$_X$ can serve as a high-quality tunneling barrier with low defect states. Therefore, a much higher TMR could possibly be achieved with a crystalline Gd$_2$O$_3$ barrier, as demonstrated in the case of MgO[48] and Al$_2$O$_3$[49]. Now that we have the temperature dependence of $G_T$ and $G_{SI}$, we are ready to compare our TMR data to that of the model. Figure 2(b) shows that our data is well described by Equation 1 and that the model is sufficient to describe the tunneling in the GdO$_X$ MTJs.

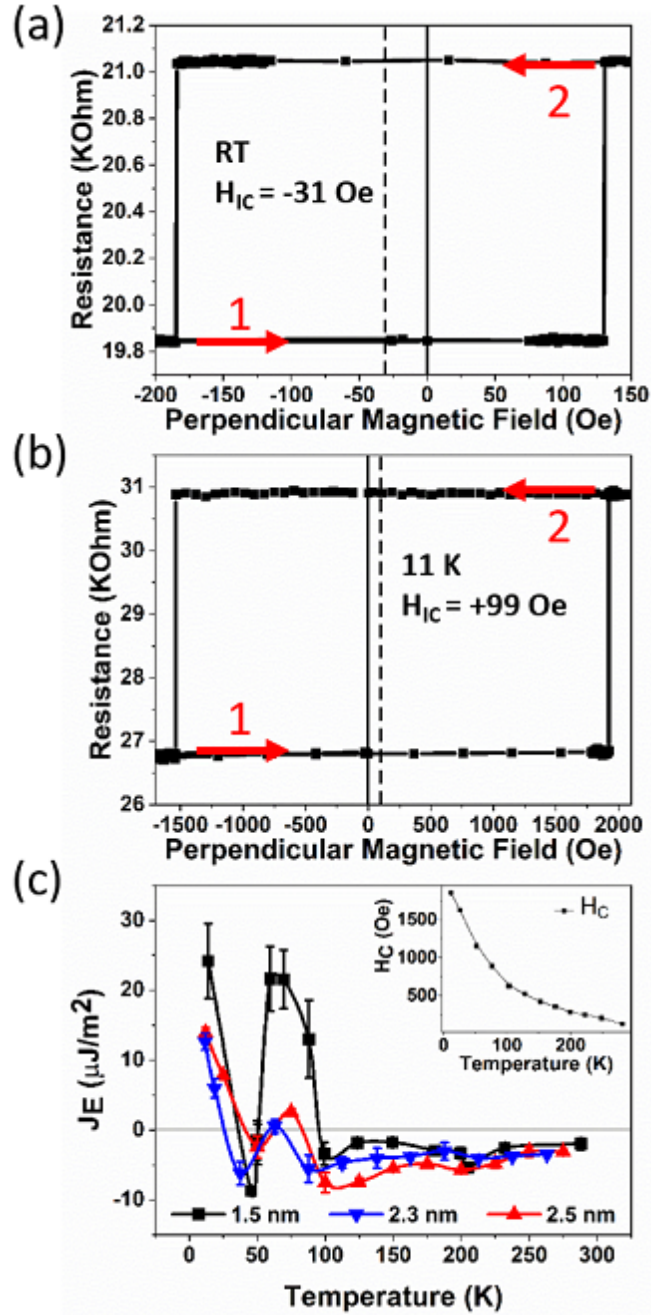

FIG. 3. (a) Minor TMR loop at RT, $H_{IC}$ = -31 Oe. (b) Minor TMR loop at 11 K, $H_{IC}$ = +99 Oe, dotted lines indicate the center of the loop for defining the $H_{IC}$ and arrows indicate the sweeping direction of the magnetic field, (c) Interlayer coupling energy vs. T showing a non-trivial temperature dependence for three thicknesses of GdO$_X$ tunneling barrier. Error bars come from averaging repeated loop measurements at each temperature. Inset. Positive switching field vs. T from RT down to 11 K showing a smooth increase for the 2.5 nm sample. The lines are a guide for the eye.

In addition to the full TMR loops, minor TMR loops were measured at each temperature by switching the soft FM layer while the hard layer magnetization remained pointing down. From the minor loop, the $H_{IC}$ can be measured from the shift of the loop along the magnetic field axis. If the minor loop is centered about a positive (negative) field, the system is said to be ferromagnetically (antiferromagnetically) coupled and prefers to be in a parallel (antiparallel) state when no magnetic field is applied. Figures 3(a) and 3(b) show the minor loops of the same $GdO_X$-pMTJ at RT and 11 K, respectively. At RT, this pMTJ shows a Hc of 157 Oe and a $H_{IC}$ of -31 Oe, consistent with the previous report[24]. Notably, when cooled to 11 K, the $H_{IC}$ changed sign to be +99 Oe, accompanied by an increase of Hc to 1730 Oe.

Previously, interlayer coupling has been studied in many systems, including spinvalves[50,51] and MTJs with both in-plane[25,28,52] and perpendicular anisotropy[53,54]. The wide range of systems in which coupling has been observed has led to several theories describing its origin in MTJs. Bruno[26] described the coupling in all metallic spinvalves based on spin-dependent reflections at the boundaries of the FM layers. He then extended his theory to insulating MTJs with the inclusion of an imaginary Fermi momentum in the barrier. According to Bruno's theory, the coupling strength across an insulating barrier should increase with temperature following a CT/sin(CT) dependence and never change sign. This temperature dependence was seen experimentally in junctions based on amorphous Si and $SiO_2$ barriers[55] but does not explain the sign change seen between Figs. 3(a) and 3(b).

In MgO based MTJs with large TMR (in-plane magnetic anisotropy), another mechanism has been shown by Katayama *et al.*[28] where the coupling is mediated through oxygen vacancies in the MgO barrier. Following the theory by Zhuravlev *et al.*[56], the oxygen vacancies act as impurity states in the barrier that mediate coupling through the so called F center resonance. This coupling is shown to be AFM if the energy level of the impurity state lies at the Fermi energy. This model predicts a smooth strengthening of the coupling with decreasing temperature, which is opposite to the Bruno model. In a subsequent study by Yang *et al.*[57], a sign change of $H_{IC}$ was predicted when the distribution of oxygen vacancies within the barrier was modified.

The interlayer coupling in pMTJs has been investigated less than in-plane MTJs, in part due to the larger difficulty involved with growing MTJs with out-of-plane easy axes. The first observation of AFM coupling in a pMTJ was made by Nistor *et al.*[53] in Co/MgO/Co junctions. Later this AFM coupling was partly explained by a model developed by Moritz *et al.*[29]. In this model, correlated roughness at the FM interfaces induces orange-peel type Neel coupling[58]. Coupling of this type is strongly dependent on the magnetic anisotropy energy. For PMA electrodes, a low anisotropy leads to FM coupling, while a larger PMA gives rise to AFM coupling. Additionally, in the AFM coupling regime, stronger PMA leads to a stronger coupling strength. An image showing the topography at the top of the $GdO_X$ layer is displayed in Fig. 4. In agreement with previous TEM findings[24], this shows a smooth layer with RMS roughness of 0.130 nm. In our previous study, a smaller $H_{IC}$ was observed when the PMA of the MTJ was increased by annealing at a higher temperature,[24] which is an indication that the AFM coupling observed in $GdO_X$-pMTJs is not described by the Moritz model. As pointed out by Nistor[30], however, the reduction of the AFM coupling after annealing at a higher temperature could be a result of reduced roughness. The measurement of $H_{IC}$ from a single sample at different temperatures provides critical information for understanding the coupling in GdO$_X$-pMTJs. According to the Moritz model, with a fixed roughness, the AFM coupling should get stronger at 11 K due to the larger PMA[39], which is not consistent with Figs. 3(a) and 3(b).

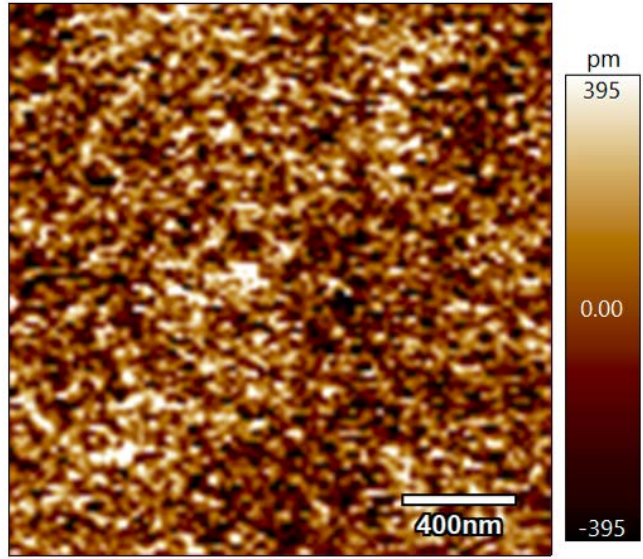

*FIG. 4. Atomic force microscopy topography image of the GdO$_X$ barrier layer.*

Figure 3(c) shows the detailed temperature dependence of the interlayer coupling energy $J_e = 1/2\, H_{IC} M_s t_m$ from RT down to 11 K for three GdO$_X$ thicknesses. Where the saturation magnetization, M$_S$ is assumed to follow the temperature dependence of spin wave excitations, $M_S = M_0(1 - \alpha T^{3/2})$ and t$_m$ is the combined thickness of the hard and soft magnetic layers. Although it is hard to tell if there is a consistent dependence of the transition temperature on the GdO$_X$ thickness with the current resolution of the data points, it is clear that the temperature dependence of J$_E$ follows the same trend for all GdO$_X$ thicknesses. All thicknesses show a non-trivial dependence that is not reflected by previous models and does not match any previous experimental data from other MTJ systems. Similarly, the J$_E$ does not follow the smooth temperature dependence of the switching field as shown in the inset of Fig. 3(c), further confirming that it is not following the trend of the sample's PMA. In addition to the large magnitude and sign change shown between RT and 11 K in Figs. 3(a) and 3(b), there is a complex temperature dependence at intermediate temperatures, including multiple switches between AFM and FM coupling states, as shown in Fig. 3(c). Another possible explanation for the change in sign of H$_{IC}$ at certain temperatures could be a reversal of the hard and soft FM layers as examined by Feng et al.[59] This possibility was explored however the crossover of the switching fields was not seen in our samples. In addition, we see no temperatures at which TMR curves do not exhibit sharp switches and a well-defined antiparallel state.

This unprecedented behavior of the interlayer coupling in GdO$_X$-pMTJs may be due to the large proximity effect[60] induced magnetic moment of the Gd ions in the barrier.[24] The induced magnetic moment of the Gd is found to be antiferromagnetically coupled to the FM electrodes and exhibits two distinct switchings that correspond to the MTJ switching fields, suggesting that they are magnetically coupled to both electrodes separately[24]. If the coupling in our GdO$_X$-pMTJs is mediated through the proximity magnetism of the GdO$_X$ barrier, it could have a much more complex dependence on temperature than that predicted by earlier models. Previous studies have shown that the H$_{IC}$ becomes negligibly small for thick enough barriers, generally vanishing for barriers thicker than 1 to 2 nm[28,30,57]. The magnetic moment of the Gd ions may also explain why the coupling in our system extends to much thicker barriers than previously demonstrated. The effective barrier thickness may be significantly smaller than the thickness of the GdO$_X$ if the proximity effect magnetizes a significant portion of each side of the barrier. A polarized neutron study is underway to investigate the thickness dependence of this proximity induced magnetism.

In summary, we have investigated the temperature dependence of GdO$_X$ based pMTJs from RT down to 11 K. Similar to AlO$_X$ and MgO based junctions, the spin-independent conductance contributes a only small proportion to the total conduction, indicative of a high-quality barrier. The most interesting result is that the H$_{IC}$ shows a non-trivial temperature dependence, including changing sign below 80 K. This behavior is not characteristic of any of the existing models, suggesting that the interlayer coupling in GdO$_X$-pMTJs is mediated through the unique magnetic properties of the GdO$_X$ tunneling barrier.

## Acknowledgments

This work was supported in part by C-SPIN, one of six centers of STARnet, a Semiconductor Research Corporation program, sponsored by MARCO and DARPA, and by the National Science Foundation through ECCS-1554011 and ECCS-1607911. Work at ORNL was supported by the Division of Scientific User Facilities of the Office of Basic Energy Sciences (BES), U.S. Department of Energy (DOE). All Cypher AFM images and data were collected in the W.M. Keck Center for Nano-Scale Imaging in the Department of Chemistry and Biochemistry at the University of Arizona. This material is based upon work supported by the National Science Foundation under Grant Number 1337371.